\documentclass[conference]{IEEEtran}

\IEEEoverridecommandlockouts

\usepackage{url}
\usepackage{array}
\usepackage{cancel}
\usepackage{siunitx}
\usepackage{multirow}
\usepackage{graphicx}
\usepackage{textcomp}
\usepackage{multirow}
\usepackage{stfloats}
\usepackage{listings}
\usepackage{colortbl}
\usepackage{algorithm}
\usepackage[T1]{fontenc}
\usepackage[10pt]{moresize}
\usepackage[utf8]{inputenc}
\usepackage[noadjust]{cite}
\usepackage{mathtools, cuted}
\usepackage[noend]{algpseudocode}
\usepackage[table,x11names]{xcolor}
\usepackage{amsmath,bm,amssymb,amsfonts}

\def\BibTeX{{\rm B\kern-.05em{\sc i\kern-.025em b}\kern-.08em
		T\kern-.1667em\lower.7ex\hbox{E}\kern-.125emX}
}

\graphicspath{{./figs/}}

\begin{document}
	
	\title{KRATT: QBF-Assisted Removal and Structural Analysis Attack Against Logic Locking}
	
	\author{
		\IEEEauthorblockN{Levent~Aksoy\IEEEauthorrefmark{2}, Muhammad Yasin\IEEEauthorrefmark{3} and Samuel~Pagliarini\IEEEauthorrefmark{2}\IEEEauthorrefmark{4}}
		\IEEEauthorblockA{\IEEEauthorrefmark{2}Department of Computer Systems, Tallinn University of Technology, Tallinn, Estonia\\
		\IEEEauthorrefmark{3}Department of Computer and Software Engineering, National University of Sciences and Technology, Islamabad, Pakistan\\
		\IEEEauthorrefmark{4}Department of Electrical and Computer Engineering, Carnegie Mellon University, Pittsburgh, USA\\
			Email: levent.aksoy@taltech.ee, m.yasin@ceme.nust.edu.pk, pagliarini@cmu.edu}
	}
	
	\maketitle
	
	\begin{abstract}
		This paper introduces KRATT, a removal and structural analysis attack against state-of-the-art logic locking techniques, such as single and double flip locking techniques (SFLTs and DFLTs). KRATT utilizes powerful quantified Boolean formulas (QBFs), which have not found widespread use in hardware security, to find the secret key of SFLTs for the first time. It can handle locked circuits under both oracle-less (OL) and oracle-guided (OG) threat models. It modifies the locked circuit and uses a prominent OL attack to make a strong guess under the OL threat model. It uses a structural analysis technique to identify promising protected input patterns and explores them using the oracle under the OG model. Experimental results on ISCAS'85, ITC'99, and HeLLO: CTF'22 benchmarks show that KRATT can break SFLTs using a QBF formulation in less than a minute, can decipher a large number of key inputs of SFLTs and DFLTs with high accuracy under the OL threat model, and can easily find the secret key of DFLTs under the OG threat model. It is shown that KRATT outperforms publicly available OL and OG attacks in terms of solution quality and run-time.
	\end{abstract}
	
	
	\section{Introduction}

In the globalized semiconductor industry, fabless design houses outsource the fabrication of their integrated circuits (ICs) to foundries, giving rise to security threats in the case of an untrusted foundry, such as piracy, overproduction, and reverse engineering, which harm the semiconductor industry financially and may even undermine national security~\cite{DSBTF15}. Many techniques, such as watermarking, digital rights management, metering, and logic locking~\cite{roy08}, have been introduced for protection against these security threats. Among these techniques, logic locking, which inserts additional logic with key inputs into the original design, has been a promising solution to many security threats. It ensures that the locked design behaves the same as the original one only when the secret key is provided. Otherwise, it generates a wrong output.

\begin{figure}[t]
	\centerline{\includegraphics[width=8.5cm]{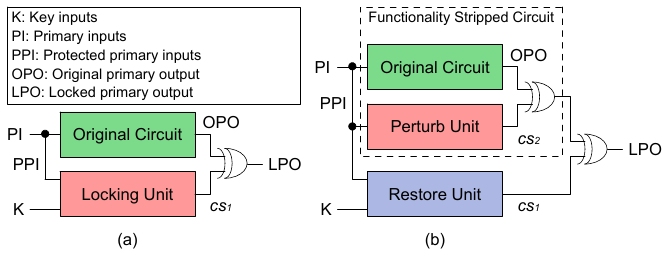}}
	\vspace*{-4mm}
	\caption{State-of-the-art logic locking techniques: (a)~SFLT; (b)~DFLT.}
	\label{fig:sdflt}
	\vspace*{-6mm}
\end{figure}

In logic locking, there are generally two main attack scenarios: (i)~in the oracle-less (OL) threat model, the adversary has only the locked netlist obtained either by reverse-engineering the layout at the untrusted foundry delivered by the design house or by reverse-engineering the functional IC obtained from the market; (ii)~in the oracle-guided (OG) threat model, the adversary also has the functional IC, which can be used as an oracle to apply inputs and observe outputs. An important milestone in logic locking is the OG satisfiability \mbox{(SAT)-based} attack~\cite{subramanyan15}, which broke all the logic locking techniques existing at that time. The SAT-based attack iteratively finds distinguishing input patterns (DIPs), which eliminate wrong keys. Thus, the state-of-the-art logic locking techniques aimed to increase the \mbox{run-time} for iterations and/or the number of iterations in the SAT-based attack~\cite{yasin16, xie19, shakya19, zhou21, ttlock, yasin17, sengupta20, kaveh19, rcalut}. As shown in Fig.~\ref{fig:sdflt}, they are generally grouped into two categories: single flip or double flip locking techniques (SFLTs and DFLTs). SFLTs~\cite{yasin16, xie19, shakya19, zhou21} use a single critical signal $cs_1$, which corrupts the original circuit for wrong keys. DFLTs~\cite{ttlock, yasin17, sengupta20, kaveh19, rcalut} use a critical signal $cs_2$ to corrupt the original design for a specific input and use a critical signal $cs_1$ to correct this corruption.

Over the years, many efficient attacks have been proposed against these state-of-the-art techniques under the OL and OG threat models~\cite{shen17,shamsi17,xu17,li19,zhang19,alaql21,sirone19,yang19,zhaokun21,sengupta21,patnaik22,limaye22,yasin20}. However, they consider either OG or OL threat model~\cite{shen17,shamsi17,xu17,li19,zhang19,alaql21}, target particular locking techniques~\cite{sirone19, yang19, zhaokun21,sengupta21}, and depend on commercial tools~\cite{limaye22,patnaik22}. Moreover, the removal attack of~\cite{yasin20} can obtain the original circuit by removing the locking unit of SFLTs from the locked circuit. However, in different scenarios, e.g., when there is no possibility to sell/fabricate the original circuit or the objective of the adversary is overproduction, finding the secret key is more valuable than obtaining the original circuit.

In this paper, we introduce a removal and structural analysis attack under both OL and OG threat models, called KRATT, developed not only for a specific logic locking technique but for a large number of \mbox{SAT-resilient} SFLTs and DFLTs. Importantly, KRATT does not rely on commercial tools. The main contributions of this paper are three-fold: (i)~it represents the problem of finding the secret key of SFLTs as a quantified Boolean formula (QBF) problem; (ii)~it shows an efficient way of determining the values of the secret key of SFLTs and DFLTs with high accuracy under the OL threat model using a circuit modification technique and the prominent OL attack SCOPE~\cite{alaql21}; (iii)~it introduces a novel approach of breaking DFLTs using structural analysis under the OG threat model. Although the use of QBF in logic locking has been hypothesized~\cite{sirone19}, to the best of our knowledge, it has never been used in breaking logic locking techniques. The manipulation of the locked circuit for the sake of an OL attack has also not been considered before. Although finding the traces of protected primary inputs has been considered~\cite{sirone19, zhaokun21}, finding them using a SAT formulation has not been proposed. Experimental results on a comprehensive set of locked circuits show that KRATT can easily break SFLTs~\cite{yasin16, xie19, shakya19, zhou21} under the OL threat model, where the QBF formulation can lead to the secret key of SFLTs~\cite{yasin16, xie19, shakya19}. It can decipher a large number of key inputs of DFLTs~\cite{yasin17, sengupta20} with high accuracy under the OL threat model. It can also break DFLTs~\cite{yasin17, kaveh19} using a little effort under the OG threat model. Although KRATT can handle a large number of locking techniques, there are still challenging techniques~\cite{yasin17,rcalut}, for which it is hard to find the secret key. Although they are out of the scope of KRATT, we describe how techniques of KRATT can be used to construct the original circuit for those techniques.


The rest of this paper is organized as follows: The background concepts are given in Section~\ref{sec:background}. KRATT is described in Section~\ref{sec:attack} and experimental results are given in Section~\ref{sec:results}. Section~\ref{sec:discussion} discusses KRATT on challenging logic locking techniques and finally, Section~\ref{sec:conclusions} concludes the paper.

	\section{Background}
\label{sec:background}

\begin{figure}[t]
	\centerline{\includegraphics[width=4.5cm]{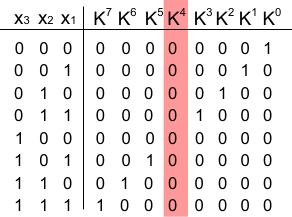}}
	\vspace*{-3mm}
	\caption{Behavior of a Boolean function locked by a point function.}
	\label{fig:pf}
	\vspace*{-6mm}
\end{figure}

\subsection{Preliminaries}

A \textit{Boolean logic function}, \(\varphi:\mathcal{B}^n \rightarrow \mathcal{B}\), where $\mathcal{B} = \{0,1\}$, over \(n\) variables \(x_1, \ldots, x_n\) maps each truth assignment to 0 or 1. The logic function \(\varphi\) in \textit{sum of products} (SOP) form, aka \textit{disjunctive normal form} (DNF), is a disjunction of \(r\) products \(p_1, \ldots, p_r\), where a \textit{product} \mbox{\(p_i = l_1 \cdot l_2 \cdot \ldots \cdot l_j\)}, \(i \leq r\) and \(j \leq n\), is a conjunction of literals. A \textit{literal} \(l_k\), \(k \leq n\), is either a variable \(x_k\) or its complement \(\overline{x_k}\). A \textit{minterm} is a product including a literal for each variable, i.e., $j=n$. An \textit{implicant} in SOP form is also a product if and only if it evaluates \(f\) to 1. Similarly, \(\varphi\) in \textit{product of sums} (POS) form, aka \textit{conjunctive normal form} (CNF), on \(n\) variables is a conjunction of \(t\) sums \(s_1, \ldots, s_t\), where a \textit{sum}, \mbox{\(s_i = l_1 + l_2 + \ldots \cdot l_j\)}, \(i \leq t\) and \mbox{\(j \leq n\),} is a disjunction of literals. A \textit{maxterm} is a sum including a literal for each variable, i.e., $j=n$. An \textit{implicant} in POS form is a sum if and only if it evaluates \(f\) to 0.


The \textit{SAT problem} is to find an assignment to the variables of a function $\varphi$ in CNF that makes \(\varphi\) to be equal to 1 or to prove that \(\varphi\) is equal to 0. The \textit{QBF problem} is the generalization of the SAT problem, in which both existential ($\exists$) and universal ($\forall$) quantifiers can be applied to each variable. 


\subsection{Related Work}

Before the SAT-based attack, earlier work focused on different types of key gates, such as \mbox{look-up} tables (LUTs), while considering the hardware complexity trade-offs~\cite{dupuis19}. After the SAT-based attack, among many other techniques, the point function has been used to generate \mbox{SAT-resilient} locked circuits~\cite{dupuis19}. Note that \textit{one-point function} evaluates to 1 at exactly one input pattern. For example, consider a Boolean function with 3 variables and suppose that it is locked by 3 key inputs using a one-point function. Fig.~\ref{fig:pf} presents its behavior under all possible keys, where $K^i$ stands for the assignment of the value $i$ in binary to key inputs, i.e., $k_3k_2k_1 = (i)_{bin}$ with $0 \leq i \leq 2^3-1$, and the logic 0 (1) value under each key denotes that the locked function is (not) equal to the original one. Note that $k_3k_2k_1 = 100$ is the secret key for our example. Since the output under each wrong key differs for one input pattern, a DIP found by the SAT-based attack eliminates only one wrong key, forcing an exponential increase in the number of DIPs required to find the secret key.

\begin{figure}[t]
	\centerline{\includegraphics[width=9.0cm]{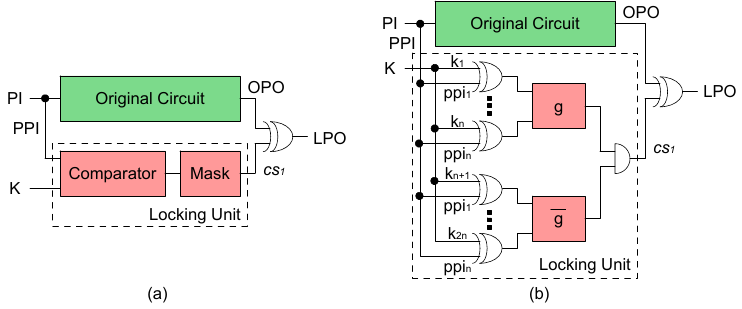}}
	\vspace*{-4mm}
	\caption{(a)~SARLock~\cite{yasin16}; (b)~AntiSAT~\cite{xie19}.}
	\label{fig:sflts}
	\vspace*{-6mm}
\end{figure}

Under the category of SFLTs, SARLock~\cite{yasin16} adds a comparator and a masking circuit connected with the original netlist in a way that it generates corruption on a specific protected primary input as shown in Fig.~\ref{fig:sflts}(a). Anti-SAT~\cite{xie19} utilizes complementary functions, which are generally composed of an {\sc and} gate tree, whose output is merged with the original circuit as shown in Fig.~\ref{fig:sflts}(b). \mbox{CAS-Lock}~\cite{shakya19} is based on the same concept of Anti-SAT, but uses a mix of {\sc and} and {\sc or} gates in the tree. Gen-Anti-SAT~\cite{zhou21} uses \mbox{non-complementary} functions to increase output corruption. 

Under the category of DFLTs, tenacious and traceless logic locking (TTLock) technique~\cite{ttlock} initially corrupts an output based on a protected primary input in the perturb unit and then, corrects this output only when the secret key is applied in the restore unit as shown in Fig.~\ref{fig:sdflt}(b). It is improved for output corruption and resiliency in~\cite{yasin17, sengupta20}. The corrupt and correct (CAC) technique~\cite{kaveh19} flips the original primary output for the protected primary input and flips it back when the primary input is equal to the protected primary input or the secret key. Techniques that hide the functionality of the restore unit in a read-proof hardware~\cite{rph} are also given in~\cite{yasin17,rcalut}. 

The OL attacks explore patterns in the structure of a locked netlist using statistical analysis~\cite{li19,zhang19,alaql21}. For example, the SCOPE attack~\cite{alaql21} is an unsupervised constant propagation technique, which analyzes each key bit of the locked design for critical features, such as area, power dissipation, and delay, which can reveal its correct value, after it is assigned to logic 0 and 1 value. Similar to the OG \mbox{SAT-based} attack, the technique of~\cite{shen17}, called DDIP, eliminates at least 2 DIPs in a single iteration. The approximate attack of~\cite{shamsi17}, called AppSAT, aims for approximate functional recovery. For SFLTs, removal attacks that find the single critical point $cs_1$ and obtain the original circuit by removing the locking unit are proposed in~\cite{yasin20}. Note that a DFLT is resilient to removal attacks since the original circuit is combined with the perturb unit, even though its restore unit can be easily removed. For DFLTs, efficient structural attacks are presented in~\cite{sirone19, yang19, zhaokun21, sengupta21, patnaik22, limaye22}. 

	\section{Removal and Structural Analysis Attack}
\label{sec:attack}

\begin{figure}[t]
	\centerline{\includegraphics[width=6.0cm]{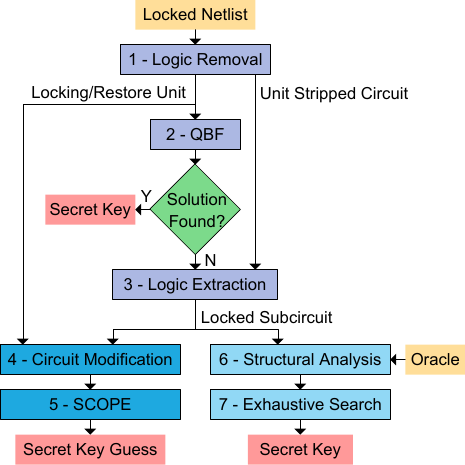}}
	\vspace*{-2mm}
	\caption{Flow of the removal and structural analysis attack.}
	\label{fig:attflow}
	\vspace*{-6mm}
\end{figure}

This section presents our removal and structural analysis attack KRATT. Its flow chart is given in Fig.~\ref{fig:attflow} and its main steps are described in detail in the following subsections.

\subsection{Logic Removal and QBF with Logic Extraction}

For both SFLTs and DFLTs under both OL and OG threat models, KRATT takes the locked netlist as input and initially extracts its locking/restore unit with the protected primary inputs and associated key inputs from the locked netlist. To do so, it follows two steps: (i)~it determines the critical signal, i.e., $cs_1$ in Fig.~\ref{fig:sdflt}, in the locked netlist by finding the output of the first gate in the paths from key inputs to primary outputs, which all the key inputs pass through; (ii)~it removes the logic cone of this critical signal and obtains the remaining of the locked netlist, called \textit{unit stripped circuit} (USC). Note that the logic shared between the locking/restore unit and USC is preserved in both circuits and the critical signal becomes another primary input of USC. In the locking/restore unit, for each protected primary input, KRATT determines its associated key input. To do so, for each protected primary input $ppi_j$, $1 \leq j \leq n$, where $n$ is the number of protected primary inputs, it finds a logic gate, whose inputs are $ppi_j$, its associated key input, or their complements. Note that a protected primary input is associated with two key inputs in Anti-SAT and its variants as shown in Fig.~\ref{fig:sflts}(b).

\begin{figure}[t]
	\centerline{\includegraphics[width=9.0cm]{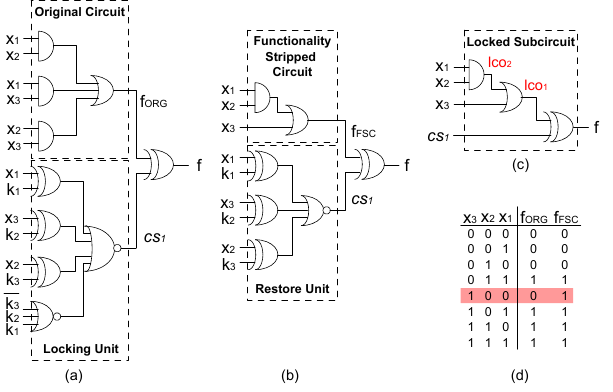}}
	\vspace*{-4mm}
	\caption{Locked majority circuits: (a)~SARLock; (b)~TTLock; (c)~locked subcircuit in TTLock; (d)~truth table for original and functionality stripped circuits in TTLock.}
	\label{fig:majlock}
	\vspace*{-6mm}
\end{figure}

Then, KRATT checks if there exist values of key inputs that set the output of the locking/restore unit, i.e., $cs_1$ in Fig.~\ref{fig:sdflt}, to a constant logic value, i.e., 0 or 1, for all possible protected primary inputs. It formalizes this problem as a QBF problem in two steps: (i)~it generates the CNF formula of this unit, $\varphi(PPI, K)$, as the conjunction of CNF formulas of each gate; (ii)~it generates two QBF problems, $\exists K\;\forall PPI,\;\varphi(PPI, K)_{cs_1=0}$ and $\exists K\;\forall PPI,\;\varphi(PPI, K)_{cs_1=1}$. Then, it solves these problems using a QBF solver. If there exists a solution to one of these QBF problems, the values of key inputs are determined to be the secret key. Note that two QBF problems are generated in order to check all possible values for the critical signal $cs_1$. 

If there exist no solutions to both QBF problems, for classification purposes, KRATT checks if the locking/restore unit realizes a comparator logic or its complement between the protected primary inputs and their associated key inputs using a SAT formulation to ensure that this unit is actually the restore unit of DFLTs. Then, it applies the logic extraction method, which takes the USC as an input and generates the \textit{locked subcircuit} including only the locked primary outputs. To do so, it finds the primary outputs reached by the critical signal $cs_1$ in USC and generates their logic cones.

As an example, consider the majority circuit locked by SARLock and TTLock using 3 key inputs as shown in Fig~\ref{fig:majlock}. Note that protected primary inputs $x_1$, $x_2$, and $x_3$ are associated with key inputs $k_1$, $k_3$, and $k_2$, respectively. The locking unit in Fig.~\ref{fig:majlock}(a) always generates logic 0 for all possible $x_1$, $x_2$, and $x_3$ values due to the 3-input {\sc nor} gate when $k_3k_2k_1 = 100$, which is the secret key found by the QBF formulation. However, since the restore unit in Fig.~\ref{fig:majlock}(b) only compares protected primary inputs with associated key inputs, there exist no solutions to the QBF problems. The subcircuit locked by TTLock is shown in Fig.~\ref{fig:majlock}(c). For the locked netlists, whose secret key cannot be found by the QBF formulation, the steps of KRATT under the OL and OG threat models are described in Sections~\ref{subsec:olatt} and~\ref{subsec:ogatt}, respectively.


\subsection{OL Attack: Circuit Modification and SCOPE}
\label{subsec:olatt}

Under the OL threat model, KRATT modifies the locked netlist to enable SCOPE~\cite{alaql21}, which may fail to make a guess or make a random guess as a standalone attack as shown in Section~\ref{sec:results}, to make a strong guess. For SFLTs, it focuses on the locking unit that includes key inputs. For Anti-SAT and its variants, where each protected primary input is associated with two key inputs, it removes all the protected primary inputs from the locking unit by setting them to a constant logic value since these inputs are not relevant to the complementary/non-complementary functions. For DFLTs, it focuses on the locked subcircuit and replaces the protected primary inputs with their associated key inputs since the information on the values of the protected primary input, although not complete, is inside the locked subcircuit. Then, it runs SCOPE on these circuits. Note that Steps 1-5 in Fig.~\ref{fig:attflow} are the steps of KRATT under the OL threat model.


\subsection{OG Attack: Structural Analysis and Exhaustive Search}
\label{subsec:ogatt}

The \textit{functionality stripped circuit} (FSC) of DFLTs as shown in Fig.~\ref{fig:sdflt}(b) is obtained after corrupting the original circuit on the protected primary input(s). For our example in Fig.~\ref{fig:majlock}(b), it is obtained by changing a maxterm of the original function into a minterm as shown in Fig.~\ref{fig:majlock}(d). The FSC includes implicants consisting of protected primary inputs and thus, their values, i.e., the secret key, can be found when exercised with oracle. 

Under the OG threat model, KRATT initially finds all the logic cones of the locked subcircuit, where their inputs are the protected primary inputs. The output of such a cone is denoted as $lco$ as shown in Fig.~\ref{fig:majlock}(c). Then, for each logic cone, it generates two sets of values of all protected primary inputs, which initially have unspecified values denoted as $X$. It determines their values by finding the input values of the logic cone when its output, i.e., $lco_i$, $1 \leq i \leq k$, where $k$ is the number of such logic cones, is set to logic 0 and 1. It formalizes this problem as a SAT problem. The reason behind setting the output of the logic cone to 0 and 1 is to try both a maxterm and minterm in the logic cone and the reason behind finding only two sets rather than all possible implicants of the logic cone is to try a small number of promising ones, as other gates in the logic cone will also be considered. For our example in Fig.~\ref{fig:majlock}(c), there exist 4 possible sets, $x_3x_2x_1 = 000$ and $x_3x_2x_1 = 100$ found for $lco_1$ when it is set to 0 and 1, respectively and $x_3x_2x_1 = X00$ and $x_3x_2x_1 = X11$ found for $lco_2$ when it is set to 0 and 1, respectively. These sets are augmented by those, if not available, where a single protected primary input is set to a constant value and all others are set to $X$ to cover all possible values, e.g., $x_3x_2x_1 = 0XX$.


Then, all these sets of values for the protected primary inputs are sorted based on the number of unspecified values in ascending order. Starting with the one including the maximum number of specified protected primary input values, for each set, KRATT generates all possible protected primary input values by exercising logic 0 and 1 values on the unspecified entries, applies it to the oracle while other primary inputs are set to logic 0, and obtains the oracle output. Then, it applies these values of primary inputs of the original design to primary inputs of the locked netlist while the values of key inputs are set to those of associated protected primary inputs and obtains the locked netlist output. If outputs of the oracle and locked netlist match as shown in Fig.~\ref{fig:pf}, it determines the secret key based on the association between the protected primary inputs and key inputs. For our example in Fig.~\ref{fig:majlock}(b), the secret key is found as $k_3k_2k_1 = 010$ when it is observed that the original design and locked netlist generate the same output at $x_3x_2x_1 = 100$, which was deduced from the logic cone with the output $lco_1$. Note that Steps 1-3 and 6-7 in Fig.~\ref{fig:attflow} are the steps of KRATT under the OG threat model.

KRATT is developed in Perl and is freely available~\cite{kratt}. It is only equipped with the QBF solver DepQBF~\cite{depqbf} and the SAT solver cryptominisat~\cite{cryptominisat}. 

	\section{Experimental Results}
\label{sec:results}

\begin{table}[t]
	\centering
	\caption{Details of the ISCAS'85 and ITC'99 circuits.}
	\vspace{-3mm}
	\begin{tabular}{|l|c|c|c|c|}
		\hline
		Circuit & \#inputs & \#outputs & \#gates & \#key inputs\\ 
		\hline \hline
		c2670  & 157 & 64  & 1193  & 64 \\
		c5315  & 178 & 123 & 2307  & 64 \\
		c6288  & 32  & 32  & 2416  & 32 \\
		b14\_C & 277 & 299 & 9768  & 128 \\
		b15\_C & 485 & 519 & 8367  & 128 \\
		b20\_C & 522 & 512 & 19683 & 128 \\
		\hline
	\end{tabular}
	\label{tab:bench}
	\vspace{-6mm}
\end{table}

\begin{table*}[t]
	\centering
	\footnotesize
	\caption{Results of OL Attacks on Locked ISCAS'85 and ITC'99 Circuits.}
	\vspace{-3mm}
	\begin{tabular}{|@{\hskip3pt}l@{\hskip3pt}|c@{\hskip3pt}|@{\hskip3pt}c@{\hskip3pt}|c@{\hskip3pt}|@{\hskip3pt}c@{\hskip3pt}|c@{\hskip3pt}|@{\hskip3pt}c@{\hskip3pt}|c@{\hskip3pt}|@{\hskip3pt}c@{\hskip3pt}|c@{\hskip3pt}|@{\hskip3pt}c@{\hskip3pt}|c@{\hskip3pt}|@{\hskip3pt}c@{\hskip3pt}|c@{\hskip3pt}|@{\hskip3pt}c@{\hskip3pt}|c|@{\hskip3pt}c@{\hskip3pt}|}
		\hline
		\multirow{4}{*}{Circuit} & \multicolumn{8}{c|}{SFLT} & \multicolumn{8}{c|}{DFLT} \\
		\cline{2-17}
		& \multicolumn{4}{c|}{Anti-SAT} & \multicolumn{4}{c|}{SARLock} & \multicolumn{4}{c|}{CAC} & \multicolumn{4}{c|}{TTLock} \\ 
		\cline{2-17}
		& \multicolumn{2}{c|}{SCOPE} & \multicolumn{2}{c|}{KRATT} & \multicolumn{2}{c|}{SCOPE} & \multicolumn{2}{c|}{KRATT} & \multicolumn{2}{c|}{SCOPE} & \multicolumn{2}{c|}{KRATT} & \multicolumn{2}{c|}{SCOPE} & \multicolumn{2}{c|}{KRATT}\\ 
		\cline{2-17}
		& cdk/dk & CPU & cdk/dk & CPU & cdk/dk & CPU & cdk/dk & CPU & cdk/dk & CPU & cdk/dk & CPU & cdk/dk & CPU & cdk/dk & CPU \\
		\hline \hline		
		c2670  & 13/23 & 3.12  & 64/64   & 0.39  & 64/64   & 3.3   & 64/64   & 0.34  & 17/26 & 3.22  & 33/64  & 64.48 & 14/26 & 3.19  & 34/64  & 64.37 \\
		c5315  & 13/22 & 3.95  & 64/64   & 0.68  & 64/64   & 3.94  & 64/64   & 0.50  & 12/19 & 3.93  & 33/64  & 64.61 & 16/31 & 4.03  & 34/64  & 64.53 \\
		c6288  & 7/12  & 2.25  & 32/32   & 0.67  & 32/32   & 2.48  & 32/32   & 0.74  & 11/18 & 2.29  & 18/32  & 64.09 & 9/14  & 2.23  & 20/32  & 63.07 \\
		b14\_C & 32/55 & 15.15 & 128/128 & 4.61  & 128/128 & 15.52 & 128/128 & 10.11 & 39/71 & 15.08 & 67/128 & 74.65 & 35/59 & 14.89 & 70/128 & 74.42 \\
		b15\_C & 22/38 & 20.04 & 128/128 & 9.01  & 128/128 & 20.42 & 128/128 & 11.91 & 18/35 & 21.41 & 64/128 & 79.57 & 43/70 & 20.22 & 68/128 & 78.70 \\
		b20\_C & 24/46 & 25.82 & 128/128 & 13.60 & 128/128 & 26.25 & 128/128 & 16.98 & 30/54 & 26.11 & 58/102 & 79.35 & 24/46 & 26.16 & 68/128 & 82.34 \\
		\hline
	\end{tabular}
	\label{tab:ol}
	\vspace{-4mm}
\end{table*}

\begin{table*}[t]
	\centering
	\footnotesize
	\caption{Results of OG Attacks on Locked ISCAS'85 and ITC'99 Circuits.}
	\vspace{-3mm}
	\begin{tabular}{|l|c@{\hskip3pt}|@{\hskip3pt}c@{\hskip3pt}|@{\hskip3pt}c@{\hskip3pt}|@{\hskip3pt}c@{\hskip3pt}|c@{\hskip3pt}|@{\hskip3pt}c@{\hskip3pt}|@{\hskip3pt}c@{\hskip3pt}|@{\hskip3pt}c@{\hskip3pt}|c@{\hskip3pt}|c@{\hskip3pt}|@{\hskip3pt}c@{\hskip3pt}|@{\hskip3pt}c@{\hskip3pt}|c@{\hskip3pt}|@{\hskip3pt}c@{\hskip3pt}|@{\hskip3pt}c@{\hskip3pt}|@{\hskip3pt}c@{\hskip3pt}|}
		\hline
		\multirow{3}{*}{Circuit} & \multicolumn{8}{c|}{SFLT} & \multicolumn{8}{c|}{DFLT} \\
		\cline{2-17}
		& \multicolumn{4}{c|}{Anti-SAT} & \multicolumn{4}{c|}{SARLock} & \multicolumn{4}{c|}{CAC} & \multicolumn{4}{c|}{TTLock} \\ 
		\cline{2-17}
		& SAT & DDIP & AppSAT & KRATT & SAT & DDIP & AppSAT & KRATT & SAT & DDIP & AppSAT & KRATT & SAT & DDIP & AppSAT & KRATT \\
		\hline \hline		
		c2670  & OoT & OoT & OoT & 0.32  & OoT & OoT & OoT & 0.33  & OoT & OoT & OoT & 70.79  & OoT & OoT & OoT & 70.50 \\
		c5315  & OoT & OoT & OoT & 0.65  & OoT & OoT & OoT & 0.47  & OoT & OoT & OoT & 76.37  & OoT & OoT & OoT & 75.94 \\
		c6288  & OoT & OoT & OoT & 0.63  & OoT & OoT & OoT & 0.70  & OoT & OoT & OoT & 163.19 & OoT & OoT & OoT & 161.21 \\
		b14\_C & OoT & OoT & OoT & 4.58  & OoT & OoT & OoT & 10.74 & OoT & OoT & OoT & 114.97 & OoT & OoT & OoT & 112.89 \\
		b15\_C & OoT & OoT & OoT & 9.14  & OoT & OoT & OoT & 11.93 & OoT & OoT & OoT & 133.30 & OoT & OoT & OoT & 131.60 \\
		b20\_C & OoT & OoT & OoT & 13.72 & OoT & OoT & OoT & 16.94 & OoT & OoT & OoT & 128.06 & OoT & OoT & OoT & 138.70 \\
		\hline
	\end{tabular}
	\label{tab:og}
	\vspace{-6mm}
\end{table*}

As the first experiment set, we used a total of six circuits from ISCAS'85 and ITC'99 benchmarks in a wide range of the number of gates. Table~\ref{tab:bench} presents the details taken from their bench files. We locked them using our implementations of Anti-SAT~\cite{xie19}, SARLock~\cite{yasin16}, CAC~\cite{kaveh19}, and TTLock~\cite{ttlock} at register transfer level (RTL) with the number of key inputs given in Table~\ref{tab:bench}. We also \textit{synthesized} the locked circuit using the Cadence Genus logic synthesis tool to break the regular structure of the locking scheme, making it harder to find the secret key, especially for removal and structural analysis attacks including KRATT. We used OL attack SCOPE~\cite{alaql21} and OG attacks, SAT-based~\cite{subramanyan15}, DDIP~\cite{shen17}, and AppSAT~\cite{shamsi17}, all of which are available to the public. The FALL attack of~\cite{sirone19}, which targets only stripped functionality logic locking (SFLL) techniques, was also run on circuits locked by TTLock, but without success. Since DDIP and AppSAT may return a wrong key in a single run, they were run multiple times with different settings. These attacks were run on a computing server including 32 Intel Xeon processing units at 3.9~GHz with 128~GB memory. The time limit for the QBF solver was set to 1 minute since finding a satisfiable solution, if exists, is generally trivial. 

\begin{figure}[t]
	\centering
	\vspace*{-6mm}
	\parbox{9.0cm}{\includegraphics[width=8.5cm]{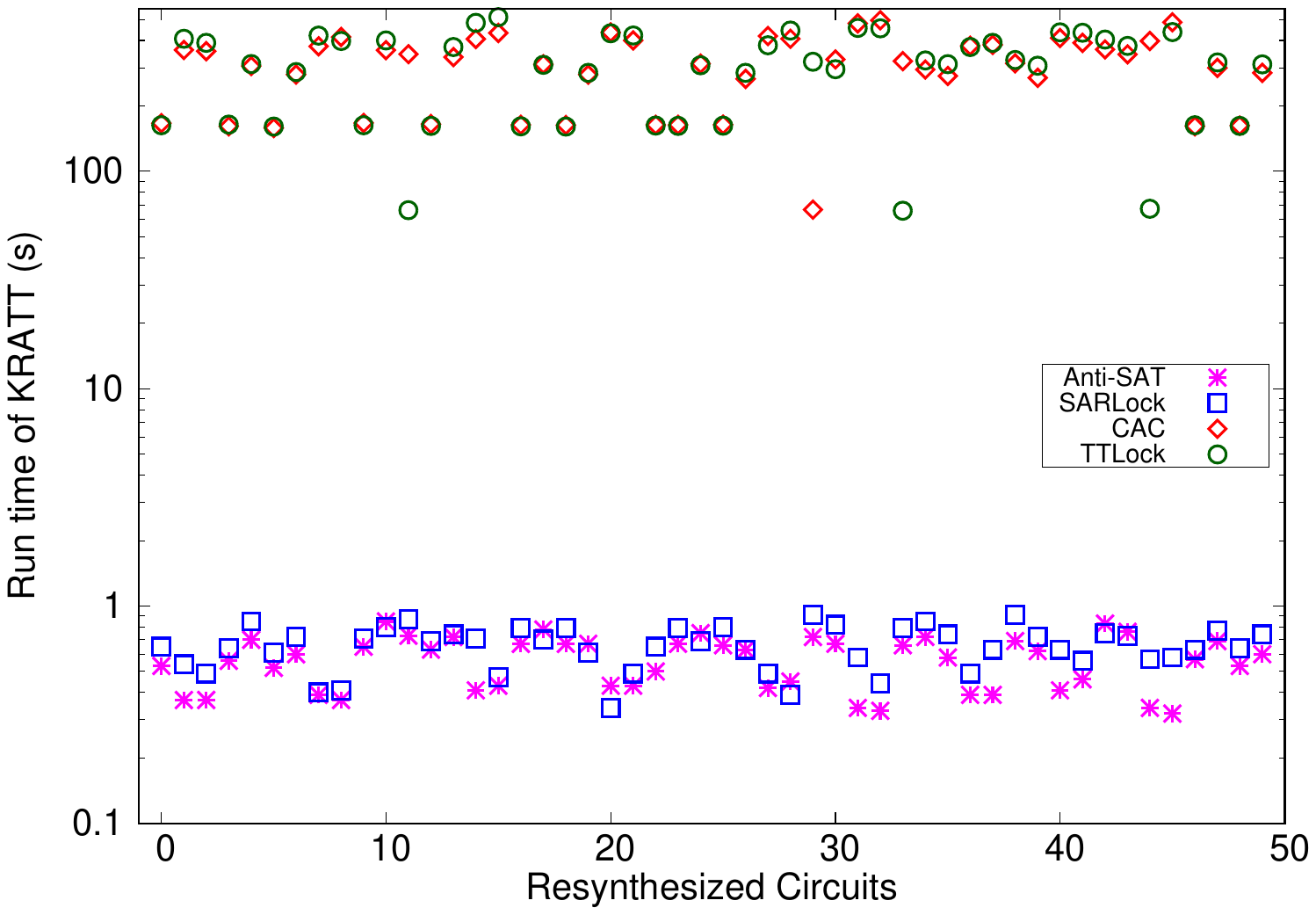}}
	\vspace*{-9mm}
	\caption{Impact of resynthesis on the run-time of KRATT.}
	\label{fig:resynth}
	\vspace*{-6mm}
\end{figure}

Tables~\ref{tab:ol} and~\ref{tab:og} present the results of OL and OG attacks, respectively. In these tables, the run-time of attacks is given in seconds. In Table~\ref{tab:ol}, $cdk$ and $dk$ are the number of correctly deciphered key inputs and deciphered key inputs, respectively. In Table~\ref{tab:og}, \textit{OoT} indicates that a solution could not be found due to the given time limit set to 2 days. 

Observe from Table~\ref{tab:ol} that while the SCOPE attack cannot decipher all key inputs, except the circuits locked by \mbox{SARLock}, KRATT can decipher all the key inputs of the locked designs, except the \textit{b20\_C} circuit locked by CAC, and can guarantee the secret key on both SFLTs. Note that KRATT finds a solution in less time than SCOPE on SFLTs since it focuses on the locking unit rather than the entire locked design and uses a QBF formulation. Observe from Table~\ref{tab:og} that the SAT-based attack and its variants cannot find a solution in the given time limit. Note that AppSAT and DDIP, which were previously shown to break AntiSAT and SARLock, respectively, fail simply because of the synthesis of locked designs and their large hardware complexity. However, KRATT can break these locked designs using a little computational effort. Note also that all the SFLTs were broken through the QBF formulation. As the complexity of circuits and the number of key inputs increase, the run-time of KRATT also increases. Its run-time on DFLTs is larger than that on SFLTs since it explores the values of possible protected primary inputs exhaustively in circuits locked by DFLTs after running the QBF solver. 

\begin{table}[t]
	\centering
	\footnotesize
	\caption{Results of OL Attacks on Circuits Locked by Gen-anti-SAT.}
	\vspace{-3mm}
	\begin{tabular}{|l|c|c|c|c|}
		\hline
		\multirow{2}{*}{Circuit} & \multicolumn{2}{c|}{SCOPE} & \multicolumn{2}{c|}{KRATT}\\
		\cline{2-5}
		& cdk/dk & CPU & cdk/dk & CPU \\
		\hline \hline
		b14\_C & 9/12 & 14.38 & 127/127 & 106.32 \\
		b15\_C & 0/0  & 19.53 & 128/128 & 137.20 \\
		b17\_C & 0/0  & 51.54 & 128/128 & 533.35 \\
		b20\_C & 4/4  & 25.07 & 128/128 & 170.28 \\
		b21\_C & 0/0  & 24.80 & 128/128 & 173.41 \\
		b22\_C & 4/4  & 34.49 & 128/128 & 261.95 \\
		\hline
	\end{tabular}
	\label{tab:genantisat}
	\vspace{-6mm}
\end{table}

In order to explore the impact of different circuit structures on the run-time of KRATT, we resynthesized the locked \textit{c6288} circuit using different design efforts and delay constraints and generated 50 functionally equivalent but structurally different circuits. This circuit was chosen because its netlists locked by DFLTs require the longest run-time under the OG threat model. Fig.~\ref{fig:resynth} presents the run-time of KRATT on these resynthesized circuits under the OG threat model. 

Note that all the SFLTs and DFLTs were again broken through the QBF formulation and structural analysis, respectively. Observe from Fig.~\ref{fig:resynth} that the impact of resynthesis on the run-time of KRATT on the circuits locked by SFLTs is less than that on circuits locked by DFLTs. The average (standard deviation) of these run-time results on resynthesized circuits locked by Anti-SAT, SARLock, CAC, and TTLock are computed as  0.56 (0.14), 0.65 (0.14), 306.85 (106.13), and 305.17 (121.57) in seconds and the ratio between the maximum and minimum run-time values on these locking techniques are found as 2.65, 2.67, 7.44, and 7.76, respectively. 

As the second experiment set, we used all the locked circuits from the Valkyrie repository~\cite{valkyrie}, including six ITC'99 benchmarks locked by Anti-SAT, CAS-Lock, Gen-Anti-SAT~\cite{zhou21}, and SARLock of SFLTs and CAC and TTLock of DFLTs using two different numbers of key inputs and 10 \textit{synthesized} circuits for each benchmark. Thus, there exist a total of 720 locked circuits. Note that the Valkyrie tool in the given repository works as a security diagnostic tool, providing the critical signals as shown in Fig.~\ref{fig:sdflt} rather than an attack finding the secret key. However, KRATT was able to find the secret key of all circuits locked by SFLTs and DFLTs under the OL and OG threat models, respectively. The QBF formulation led to the secret key of all the 120 circuits locked by CAS-Lock and 112 out of 120 circuits locked by SARLock, i.e., a total of 232 locked circuits. As an interesting result, Table~\ref{tab:genantisat} presents solutions of attacks under the OL threat model on a single circuit for each benchmark locked by Gen-Anti-SAT using 128 key inputs, whose secret key could not be found through the QBF formulation due to the non-complementary function.

Observe from Table~\ref{tab:genantisat} that while SCOPE can decipher a small number of key inputs on the entire locked circuit, KRATT can correctly decipher all the key inputs on the modified locking unit. Note that on \textit{b14\_C} circuit, it was proved that the secret key was found when the value of the missing key input was set to logic 0 or 1.




\begin{table*}[t]
	\centering
	\footnotesize
	\caption{Details of Locked HeLLO: CTF'22 Circuits and Results of OL and OG Attacks.}
	\vspace{-3mm}
	\begin{tabular}{|l|c|c|c|c|c|c|c|c|c|c|}
		\hline
		\multirow{3}{*}{Circuit} & \multicolumn{4}{c|}{Locked Circuit Details} & \multicolumn{4}{c|}{OL Attacks} & \multicolumn{2}{c|}{OG Attacks}\\
		\cline{2-11}
		& \multirow{2}{*}{\#inputs} & \multirow{2}{*}{\#outputs} & \multirow{2}{*}{\#gates} & \multirow{2}{*}{\#key inputs} & \multicolumn{2}{c|}{SCOPE} & \multicolumn{2}{c|}{KRATT} & \multirow{2}{*}{SAT} & \multirow{2}{*}{KRATT} \\ 
		\cline{6-9}
		&                           &                            &                          &                               & cdk/dk & CPU & cdk/dk & CPU & & \\
		\hline \hline
		final\_v1 & 767  & 757  & 17144 & 87& 0/0 & 261.19 & 73/87 & 194.61 & 1117.05  & 350.22 \\
		final\_v2 & 1452 & 1445 & 27440 & 47& 0/0 & 39.73  & 34/46 & 99.45  & OoT      & 2186.56 \\
		final\_v3 & 522  & 1    & 93    & 29& 0/0 & 1.94   & 25/29 & 62.26  & 20448.65 & 63.97 \\
		\hline
	\end{tabular}
	\label{tab:hello}
	\vspace{-6mm}
\end{table*}

As the third experiment set, we used the circuits locked by SFLL from the HeLLO: CTF'22 competition. Table~\ref{tab:hello} presents the details of the locked circuits taken from their bench files and the solutions of OL and OG attacks. We highlight that the FALL attack of~\cite{sirone19} was not successful on these circuits.

Observe from Table~\ref{tab:hello} that KRATT can decipher all the key inputs, except \textit{final\_v2}, with high accuracy under the OL threat model, where no secret key was reported to be found by the competition participants. Also, it can find the secret key of all locked circuits using less run-time than the SAT-based attack under the OG threat model. The reason KRATT takes a longer time to break the \textit{final\_v2} circuit than others is due to a large number of promising protected primary input candidates.

	\section{Discussion}
\label{sec:discussion}

In row/column-activated-LUT~\cite{rcalut} and SFLL-Flex~\cite{yasin17} techniques, the original circuit corrupted on a number of protected primary inputs is corrected by the restore unit implemented in a \mbox{read-proof} hardware~\cite{rph}. Since the restore unit is hidden from the adversary, and thus, the association of protected primary inputs with key inputs, KRATT cannot find the secret key as any other attack. However, for row-activated-LUT of~\cite{rcalut} or SFLL-Flex techniques, the structural analysis and exhaustive search of KRATT described in Section~\ref{subsec:ogatt} can be used to find all the protected primary inputs and thus, the original circuit can be constructed after adding these values into the FSC using a comparator and XOR logic. 
	\section{Conclusions}
\label{sec:conclusions}

This paper presented a novel removal and structural analysis attack, called KRATT, which can break a large number of state-of-the-art \mbox{SAT-resilient} logic locking techniques. KRATT utilizes the SAT and, most importantly, QBF formulations to find the secret key of locked circuits under both OL and OG threat models. It was shown that it can find the secret key of circuits locked by SFLTs using the QBF formulation, can decipher values of key inputs with higher accuracy under the OL threat model with respect to a prominent OL attack, and can break DFLTs under the OG threat model, where \mbox{well-known} logic locking attacks cannot find a solution. It was also shown that hardware complexity, resynthesis, and the number of key inputs have a moderate impact on its \mbox{run-time}. 

	\section{Acknowledgment}

This work was supported by the EU through the European Social Fund in the context of the project “ICT programme”.

	\bibliography{date24}
	\bibliographystyle{IEEEtran}

\end{document}